\documentclass[pre,twocolumn,aps,superscriptaddress,showpacs,floatfix]{revtex4}
\usepackage[centertags]{amsmath}
\usepackage{amssymb}
\usepackage{amsthm}
\usepackage{graphicx}
\usepackage{epsfig}

\begin{document}

\title{Kinetics of self-induced aggregation of Brownian particles:
 non-Markovian and non-Gaussian features}

\author{Pulak Kumar Ghosh$^1${\footnote{e-mail: gpulakchem@gmail.com}}, Monoj Kumar Sen$^2$ and Bidhan
Chandra Bag$^2${\footnote{Author for correspondence, e-mail:
pcbcb@rediffmail.com}}}

\affiliation{ $^1$Indian Association for the Cultivation of Science,
Jadavpur, Kolkata 700 032, India\\
$^2$Department of Chemistry, Visva-Bharati, Santiniketan 731 235,
India}

\begin{abstract}
In this paper we have studied a model for self-induced aggregation
in Brownian particle incorporating the non-Markovian and
non-Gaussian character of the associated random noise process. In
this model the time evolution of each individual is guided by an
over-damped Langevin equation of motion with a non-local drift
resulting from the local unbalance distributions of the other
individuals. Our simulation result shows that colored nose can
induce the cluster formation even at large noise strength. Another
observation is that critical noise strength grows very rapidly
with increase of noise correlation time for Gaussian noise than
non Gaussian one. However, at long time limit the cluster number
in aggregation process decreases with time following a power law.
The exponent in the power law increases remarkable for switching
from Markovian to non Markovian noise process.
\end{abstract}

\maketitle
\section{Introduction}
Formation of large spatial structure and clusters by the
aggregation of small species joining each other constitutes a
broad area of research in pure and applied
sciences\cite{pru,flo,jae,oku}. General properties of aggregation
dynamics include microphysics of clouds and
precipitation\cite{pru}, principle of polymer formation\cite{flo},
different types of ecological problems\cite{oku}, etc., to mention
a few. The most important direction of the theory of aggregation
dynamics leads us to the realm of biological systems in which
cooperative activity among individuals usually involves social
behaviours\cite{oku,fli}. Still now its underlying mechanism in
various biological systems remains unknown. A number of attempt
have been made to established rigorous and quantitative basis of
the emergence of cooperation among
individuals\cite{lev,now,nak,smi}. The kinship theory is the first
and important step in this direction which is based on genetic
arguments\cite{ham}. But it does not account correctly the
explanation of cooperation that involves among unrelated
individuals (individuals having no common genes). On the other
hand, according to some models of mathematical population biology
and game theory aggregation is the result of \emph{short} and
\emph{long} range interactions among different
individuals\cite{vic,ger}. According to these models the
aggregation dynamics of social biological systems has the same
statistical basis as that have been used in fluid dynamics and
condensed matter physics.

The essential requirements for a general and simple theoretical
description of aggregation dynamics in social communities involve
the space of state (where each point represents the status of the
individuals) and strategy\cite{ham,lot} (that is the rule
according to which the individuals players decide to change their
status in response to partial and complete information about the
action of the other players). Based on game theoretical approach,
Sigmund and Nowak\cite{now1} assumed that cooperation to work in
evolved social systems require the knowledge of \emph{reputation
or status} of their members (players). The reputation is denoted
by a dynamical coordinate, $S$ known as image score. The image
score is assigned to each player which is the identity to the
other member of community and indicates both wether an individual
provides help and if he/she is worthy of being help.  Recently,
Cecconi \emph{et al}\cite{cec1,cec2} studied the model suggested
by Sigmund and Nowak\cite{now1} to test the hypothesis of
emergence of cooperation by indirect reciprocity among unrelated
individuals. In their study\cite{cec1,cec2} they assumed nonlinear
Fokker-Planck equation for the population of individuals with a
certain image score. The equation had a nonlocal drift term that
characterized the strategy. Based on the stochastic
differential\cite{stra,gar} equations corresponding to a Brownian
dynamics with drift induced by the spatial distribution of other
random worker the authors have also studied the aggregation
dynamics\cite{cec3}. In this study they assumed that the random
fluctuations correspond to the white Gaussian noise. However, the
random fluctuations in the social communities problem are in
general non thermal in origin. They may appear as a result  of
complicated inherent dynamics and therefore the noise of non
thermal origin may be non Gaussian and correlated in
characteristics. The correlated noise may have important role in
the context of cooperation which is related to aggregation
dynamics and others since correlation among the particles
increases with increase of noise correlation time. Keeping in mind
of this aspect we have extended the study of the aggregation
dynamics of the self-induced model\cite{cec3} of the social
biological system in non-Markovian and non-Gaussian limit. Our aim
is to explore how the dynamics of aggregation to form cluster
depends on non-Markovian and non-Gaussian properties of the noise
process. Here it is to be noted that the noise of biological(non
thermal) origin is non Gaussian in character. Recent experimental
and theoretical studies in neural network and sensory
systems\cite{weis,noz} offer strong indication that the noise
sources in these systems could be non-Gaussian. The noise of
biological origin in many cases is due to nonlinear dynamics which
may be correlated and non-Gaussian in character, specifically, in
the context  biological evolution \cite{weis,newman}. The role of
colored non Gaussian noise in the barrier crossing dynamics, the
stochastic resonance and complex net work has been explored by
several authors \cite{bag}.

\section{The Model}
Consider a system consists of $N$ individuals which change their
$x_i$ state according to majority rule. $x_i$  denote the reputation
score of $i^{th}$ member or the position in a possible chemotaxis
description or some other amplitude characterizing the role of
individual within population biology framework. We assumed that each
individual changes $x_i$ by the following stochastic equation of
motion.
\begin{eqnarray}\label{2.1}
\dot{x_j} =  v(x_j)+  \eta(t)
\end{eqnarray}
$v$ denotes the drift, which is given by the following expression
\begin{eqnarray}\label{2.2}
v(x_i) = \lambda
\frac{w_{+}(x_i,t)-w_{-}(x_i,t)}{w_{+}(x_i,t)-w_{-}(x_i,t)}
\end{eqnarray}
where $w_{\pm}$ are defined by
\begin{eqnarray}\label{2.3}
w_{\pm}(x)=\sum_j \Theta
[\pm(x_j-x_i)]\exp{\left(-\alpha|x_j-x_i|\right)}
\end{eqnarray}
$\Theta$ is the unitary step function. The equations (\ref{2.1})
and (\ref{2.2}) reveal that the velocity at which an individual
decides to move to the left or to the right depends on the
difference $w_+ - w_-$. The magnitude of $w_{\pm}$ depends on the
exponential weight with a coefficient $\alpha=1/r_0$, $r_0$
specify the range up to which one individual still perceives the
presence and the influence of the other member of the group. So
according to the above description aggregation is the result from
preferably migration of individuals depending on the sign of $v$.
The term $w_+ - w_-$ in the denominator of Eq.(\ref{2.2}) present
the normalization factor and so the velocity bounded within
$[-\lambda,\lambda]$.

\begin{figure}[!htb]
\includegraphics[width = 8cm,angle=0,clip]{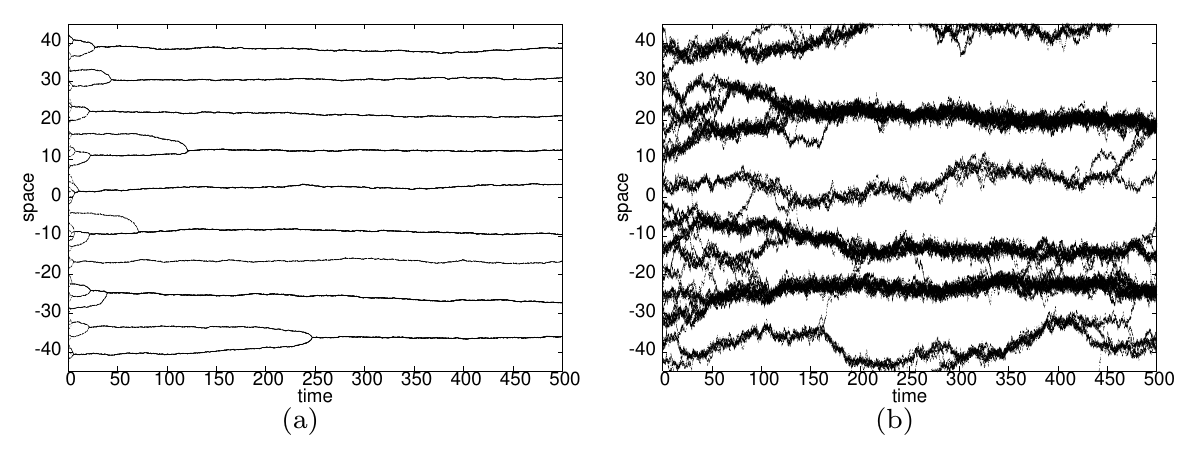}
\includegraphics[width = 8cm,angle=0,clip]{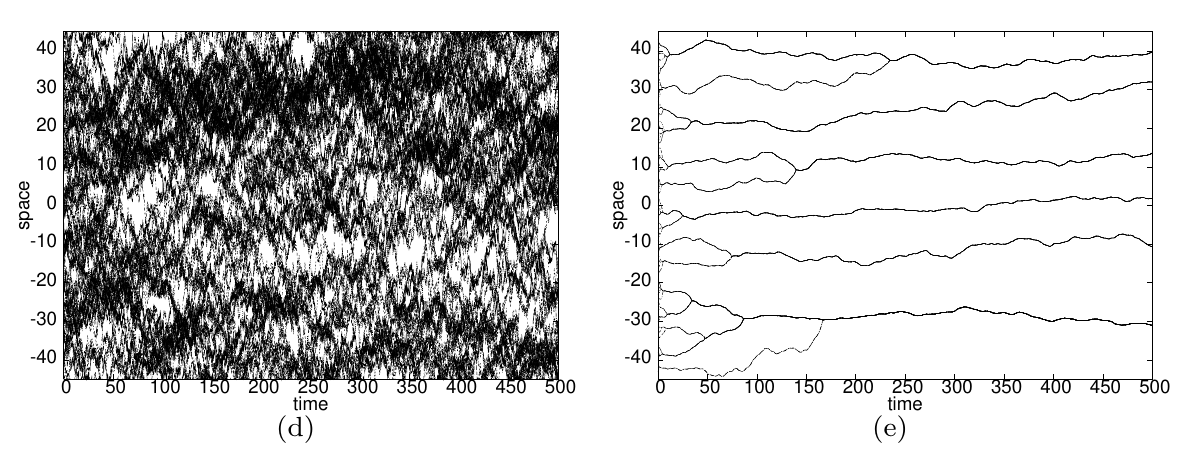}
\caption{
(Color online) This figure refer to system dynamics with 100
particles staring from uniformly distributed initial conditions in
the spacial range $[-44,+44]$. The parameter set for sub-figures
(a) $p=1,\;\tau=0.01,\; D=0.01$, (b) $p=1,\;\tau=0.01,\; D=0.5$
(c) $p=1,\;\tau=0.01,\; D=2.0$ (d) $p=1,\;\tau=5.0,\; D=0.5$ (d)
$p=1,\;\tau=8.0,\; D=2.0$ (f) $p=1,\;\tau=0.01,\; D=0.5$.
$\alpha=1$ and $\lambda=1$ for the all sub-figures} \label{fig0}
\end{figure}

We assume that $\eta(t)$  in the above equation is colored noise
which may be of Gaussian or non-Gaussian type depending on the
situation. The non-Gaussian noise can be generated from the solution
of the following Langevin equation\cite{bor11}
\begin{equation}\label{2.3}
\dot{\eta} =  -\frac{1}{\tau}\frac{d}{d\eta} V_{p}(\eta)
+\frac{\sqrt{D}}{\tau} \zeta(t)
\end{equation}
\noindent $\zeta(t)$ being a standard Gaussian noise of zero mean
and its two-time correlation given by
\begin{equation}\label{2.4}
\langle \zeta(t) \zeta(t') = 2 \delta(t-t')
\end{equation}
and
\begin{equation}\label{2.5}
V_{p}(\eta) =  \frac{D}{\tau(p-1)} \ln[1+\alpha_1(p-1)\eta^2/2)] .
\end{equation}
\noindent Here the form for noise $\eta$ allows us to control the
departure from the Gaussian behavior easily by changing a single
parameter $p$. $D$ and $\tau$ are noise parameters related to the
noise intensity and the correlation time of $\eta$. The parameter
$\alpha_1$ in Eq.(\ref{2.5}) is defined as
\begin{equation}\label{2.6}
\alpha_1= \frac{\tau}{D} \; \;.
\end{equation}
\noindent Now we consider two different situations. For $p=1$
Eq.(\ref{2.3}) becomes
\begin{equation}\label{2.7}
\dot{\eta} =  -\frac{\eta}{\tau} +\frac{\sqrt{D}}{\tau} \zeta(t) \;
\;.
\end{equation}
i. e., the time evolution equation of Ornstein-Uhlenbeck noise
process\cite{han} for which the correlation function $\langle
\eta(t) \eta(0) \rangle$ decays exponentially
\begin{equation}\label{2.8}
\langle \eta(t) \eta(0) \rangle=D/\tau^2\exp(-t/\tau) \; \;.
\end{equation}
\noindent Thus $\tau$ is the correlation time of the
Ornstein-Uhlenbeck noise. In the next step we consider a situation
where $p>1$. For $p>1$ the stationary properties of the noise
$\eta$, including the time correlation function, have been studied
in \cite{Fuen} and here we summarize the main results. The
stationary probability distribution is given by
\begin{equation}\label{2.9}
P(\eta)=\frac{1}{Z_{p}}\left[1+\alpha_1 (p-1)
\frac{\eta_i^2}{2}\right]^{\frac{-1}{p-1}} \; \; \;,
\end{equation}
\noindent where $Z_{p}$ is the normalization factor and given by
\begin{eqnarray}\label{2.10}
Z_{p}=\int_{-\infty}^{\infty}d\eta \left[1+\alpha_1 (p-1)
\frac{\eta^2}{2}\right]^{\frac{-1}{p-1}}  \nonumber\\
=  \sqrt{\frac{\pi}{\alpha_1 (p-1)}}\frac{\Gamma(1/(p-1)-1/2)}
{\Gamma(1/(p-1))} \; \; \;,
\end{eqnarray}
\noindent $\Gamma$ indicates the Gamma function. This distribution
can be normalized only for $p<3$. Since the above distribution
function is an even function of $\eta$, the first moment, $\langle
\eta \rangle$, is always equal to zero, and the second moment is
given by
\begin{equation}\label{2.11}
\langle \eta_{p}^2 \rangle  = \frac{2 D}{\tau (5-3 p)} \; \; \;,
\end{equation}
\noindent which is finite only for $p<5/3$. Furthermore, for $p<1$,
the distribution has a cut-off and it is only defined for
\begin{equation}\label{2.12}
|\eta| < \eta_{c} \equiv \sqrt{\frac{2D}{\tau (1-p)}} \; \;.
\end{equation}
\noindent Finally, the correlation time of non-Gaussian noise $\tau$
of the stationary regime of the process $\eta(t)$ diverges near
$p=5/3$ and it can be approximated over the whole range of values of
$p$ as
\begin{equation}\label{2.13}
\tau _p\simeq 2\tau/(5-3p) \; \;.
\end{equation}
\noindent Clearly, when $p_i\rightarrow 1$, we recover the limit of
$\eta_i$ being a Gaussian colored noise, the Ornstein-Uhlenbeck
process since in this limit the term in the square bracket of
Eq.(\ref{2.9}) can be written as

\begin{equation}\label{2.14}
1+\alpha_1 (p-1)
\frac{\eta^2}{2}=\exp(\alpha_1(p-1)\frac{\eta^2}{2}) \; \;,
\end{equation}
\noindent and therefore Eq.(\ref{2.9}) becomes
\begin{equation}\label{2.15}
P(\eta)=\frac{1}{Z_{1}}\exp(-\alpha_1 \eta^2/2)  \; \; \;,
\end{equation}
\noindent with
\begin{equation}\label{2.16}
Z_{1}=\sqrt{\pi/\alpha_1} \; \; \;.
\end{equation}

\noindent Here we would like to note that Eq.(\ref{2.11}) shows
that for a given external noise strength $D$ and noise correlation
time $\tau$ the variance of the non-Gaussian is higher than that
of the Gaussian noise for $p>1$, i. e
\begin{equation}\label{2.17}
\langle \eta_{p}^2 \rangle > \langle \eta^2 \rangle  \; \;.
\end{equation}
\noindent Similarly Eq.(2.14) implies that $\tau_{p} > \tau$ for
$p>1$.

\begin{figure}[!htb]
\includegraphics[width = 8cm,angle=0,clip]{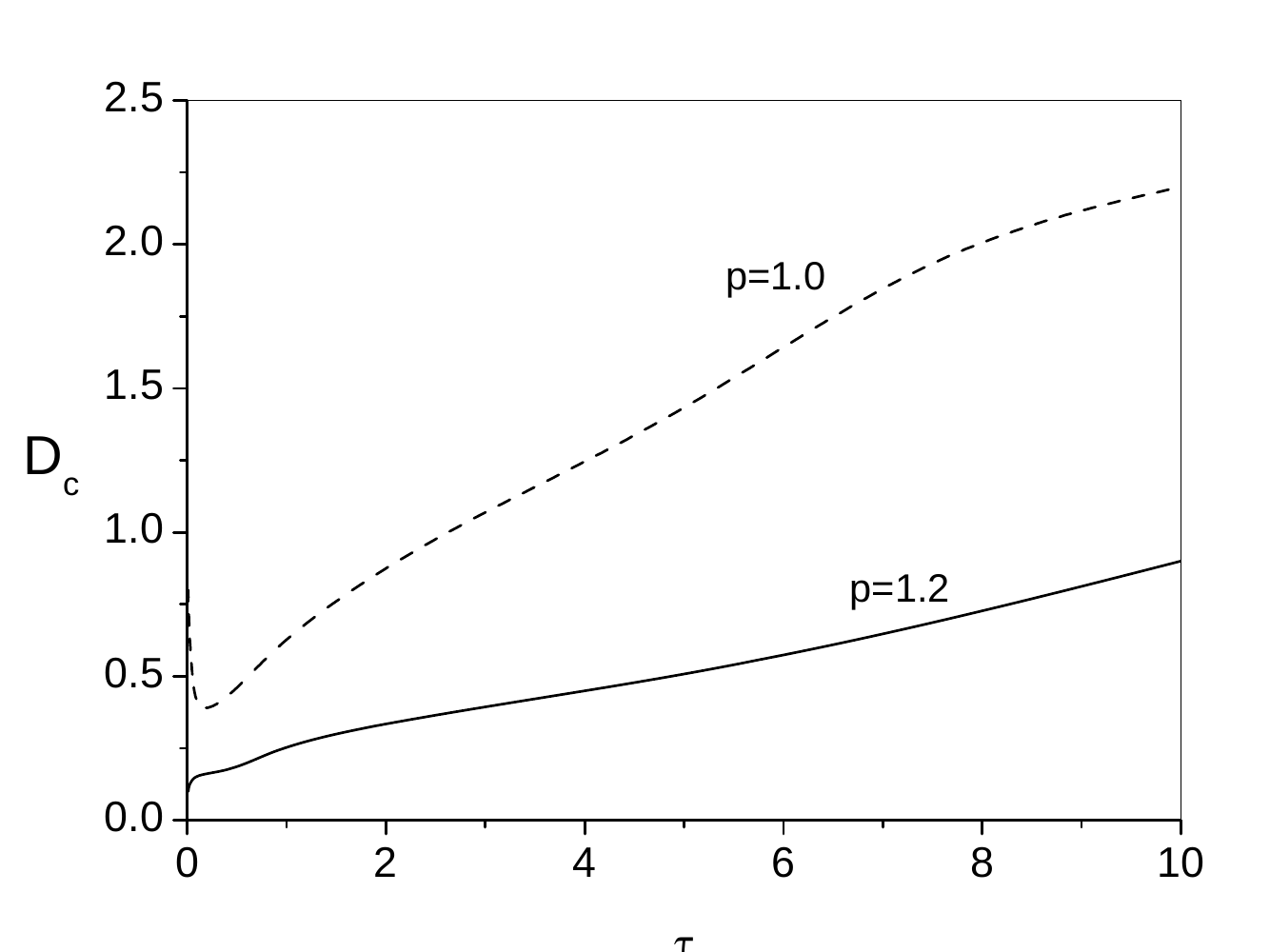} \caption{
(Color online) Number of cluster vs time plot for different values
of the noise strength to point out the critical value of noise
strength. The parameters for the plots are $\alpha=1,\; \lambda=1,
\rho=4,\;\epsilon=0.25,\;\tau=1,\;p=1$} \label{fig0}
\end{figure}

\section{Result and discussion}
Based on the above mentioned model we have numerically
investigated the aggregation dynamics. To follow the dynamics of
the each individuals present in the system we solve
 $N$ number coupled stochastic equations (Eq.(2.1)) along with the
equation for noise process (2.4) simultaneously using standard
Heun's algorithm. $N$ refers to the number of individuals present
in the system. A very small time step($\Delta t$) of $0.01$ for
numerical integration has been used. For the initial coordinates
we have assumed that at $t=0$ all the particles are uniformly
distributed in the space.
However, to exhibit the role of noise strength, noise correlation
time and other noise parameters on the cluster formation dynamics
we plot coordinates of all the particles {\it vs.} time, $t$ in
Fig.1. It shows that the cluster formation tendency as well as
size of the cluster increases as the noise strength increases up
to a critical value (Figs.1(a,b)). After the critical value of the
noise strength all the individual particles exhibits normal
Brownian motion (Fig.1c). But if we increase the noise correlation
time then even at high noise strength cluster formation is
possible. It has been shown in the Figs.1(d,e). Thus colored noise
can induce the cluster formation. Further more, Fig.1e
demonstrates that there may be a phase transition from the
clustered state to the state where particles are uniformly
distributed over the space if one switches from Gaussian to non
Gaussian noise(Fig.1f). These observations can be rationalize in
the following way. The cluster formation is a result of
cooperation among the particles. By virtue of diffusion in the
presence of noise the particle which is far apparat from the
nucleation center of the strong cluster may enter into the cluster
zone. Thus noise accelerates to form strong and bigger cluster.
But if the noise strength is very high then the diffusion
dominates over the cooperative effect and we observe the phase
where particle almost uniformly distributed over the space instead
of cluster formation. Now we come to the point how noise
correlation time can induce the cluster formation. With increase
of noise correlation time the variance of the noise decreases and
thereby diffusion of the particle suppressed as the non Markovian
nature of the noise grows. Not only that, the noise correlation
time strongly effects the drift term of the dynamics
\cite{han,pm}. The drift term in the present problem accounts the
extent of cooperation among the particles which leads to cluster
formation. Thus it is apparent in the cluster formation dynamics
that the colored noise induced nucleation is a result of the
extension of the cooperation and the correlation among the
particles as well as suppression of diffusion nature of the
particles with increase of noise correlation time. Finally, we
consider how the cluster formation is suppressed by the non
Gaussian noise. For a given noise strength the variance of the non
gaussian noise is much higher than the Gaussian noise(see Eq.()).
As a result of that for a given noise strength the diffusion may
dominates over cooperative effect for non Gaussian noise and there
is no cluster formation. But for the same noise strength the
cluster formation may be possible for Gaussian noise due to weak
diffusion compared to non Gaussian noise. Thus there may exist
phase transition phenomenon if one switches from non-Gaussian to
Gaussian  noise or vice versa.

\begin{figure}[!htb]
\includegraphics[width = 8cm,angle=0,clip]{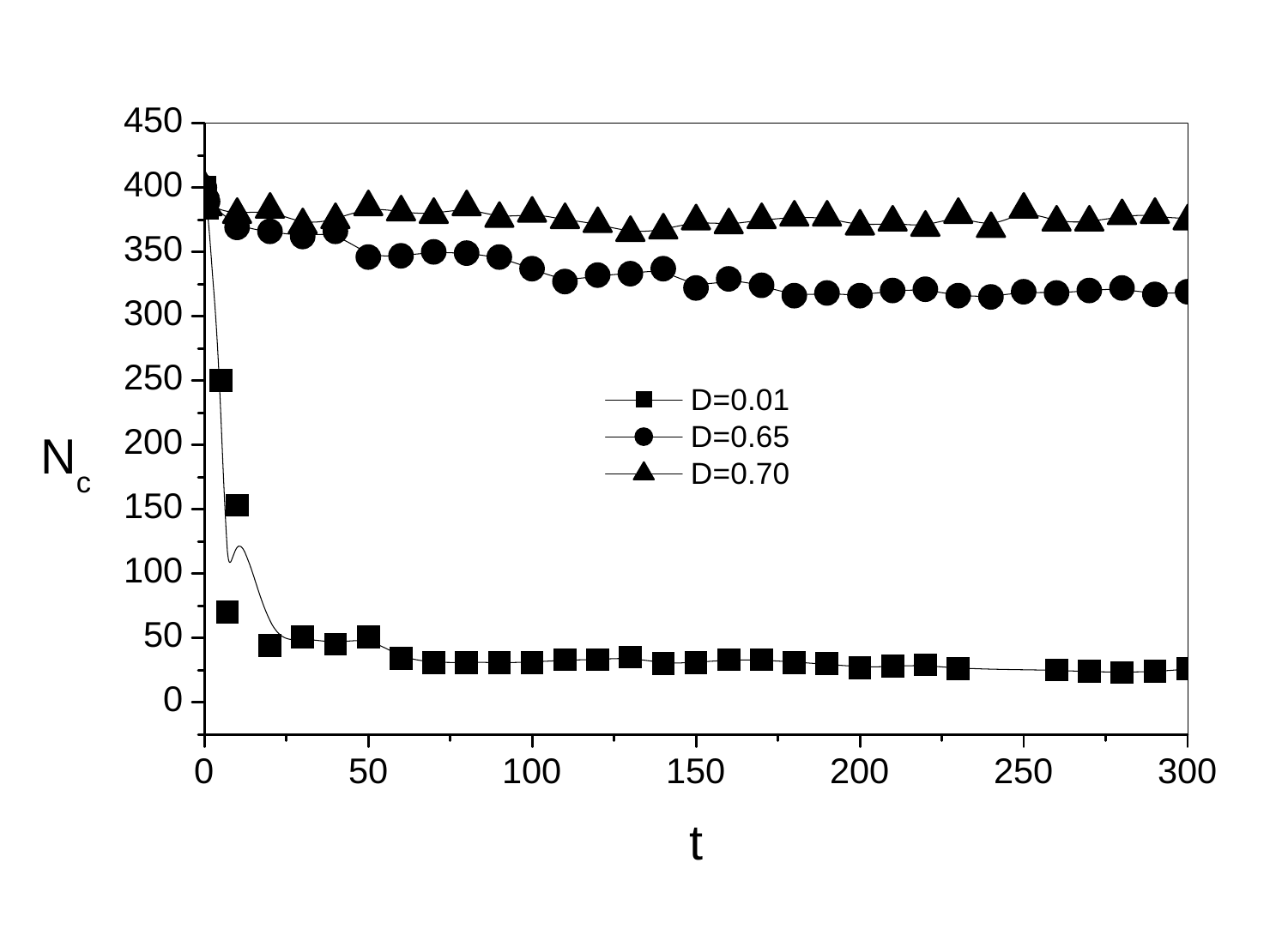} \caption{
(Color online)This figure present the critical parameter regime
for noise correlation and noise strength for different values of
the non-Gaussian parameters. Other system parameters for the plots
are $\alpha=1,\; \lambda=1$.}
\end{figure}

\begin{figure}[!htb]
\includegraphics[width = 8cm,angle=0,clip]{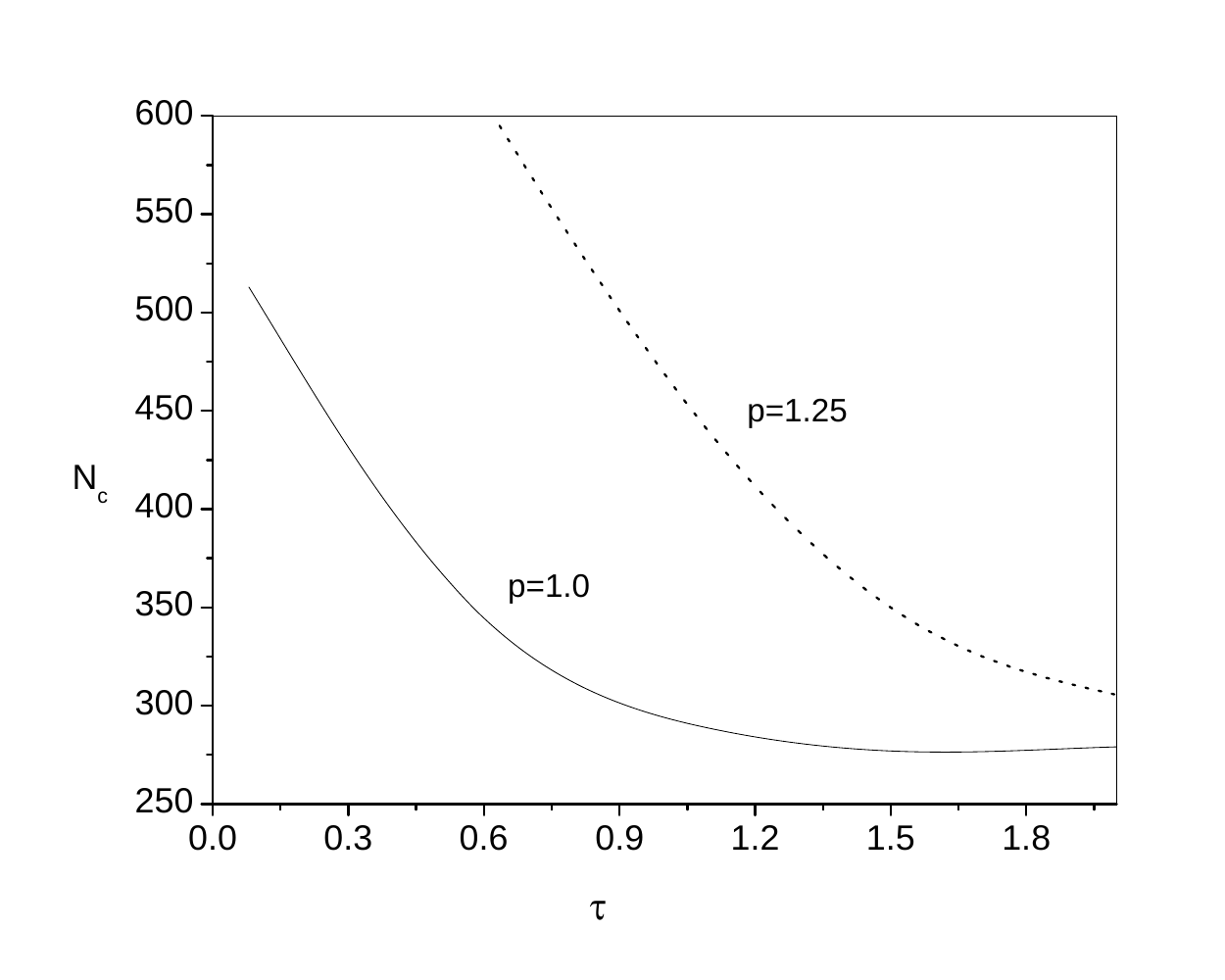} \caption{
 This plot present the variation of the cluster
number as a function of the correlation time of the noise for
different values of the non-Gaussian parameter(p). The other
parameter sets are $\lambda=1,\; \alpha=1,\; \rho=4,\;
\epsilon=0.25$.}
\end{figure}


\noindent The above discussion implies that there is a critical
noise strength($D_c$) above which cluster formation is not
possible for the given parameter set(Figs. 1(a)-1(c)). One can
determine it numerically from the plot of cluster (it is defined
as a set of particles within a given cutoff distance $\epsilon$.
$\epsilon$ is called resolution.) number($N_c$) {\it vs.} time
(t). To demonstrate this we have plotted $N_c$ {\it vs.} $t$ in
Fig.2 for the particle density $\rho=4.0$. It shows that for the
given parameter set for the Fig.3 the $D_c$ is very much close to
$0.70$ since cluster number remains almost close to its initial
value. However, to explore its dependence on the noise correlation
time we have plotted $D_c$ {\it vs.} noise correlation time in
Fig.3. It exhibits that the value of $D_c$ changes rapidly for the
Gaussian noise than the non Gaussian one. This is because of the
higher variance for former than later for the given noise strength
$D$. For the same reason the value of $D_c$ in general is higher
for Gaussian noise than that of non Gaussian noise. It is to be
noted here that the Fig.3 is a nice demonstration regarding the
dependence of phase transition on the noise strength($D$) and the
noise correlation time ($\tau$). The parameter regime below the
dashed curve is corresponding to the clustered phase for the
Gaussian noise. Similarly the points below the solid curve
represents this phase for the non Gaussian noise.

In the next step we have investigated the variation of cluster
number at stationary state with noise correlation time($\tau$) and
plotted in Fig.4. It exhibits that the cluster number rapidly
falls with $\tau$ for the Gaussian noise than the non Gaussian
noise. As the correlation as well as the cooperation among the
particles increases with increase of $tau$ the strong and bigger
cluster is formed for larger noise correlation time. Because of
that the cluster number reduces for both the Gaussian and non
Gaussian noises as tow grows. The slow decrease of cluster number
for non Gaussian noise compared to Gaussian noise is due to higher
noise variance (after certain value of noise variance it is
difficult to form cluster) for former than later.

\begin{figure}[!htb]
\includegraphics[width = 8cm,angle=0,clip]{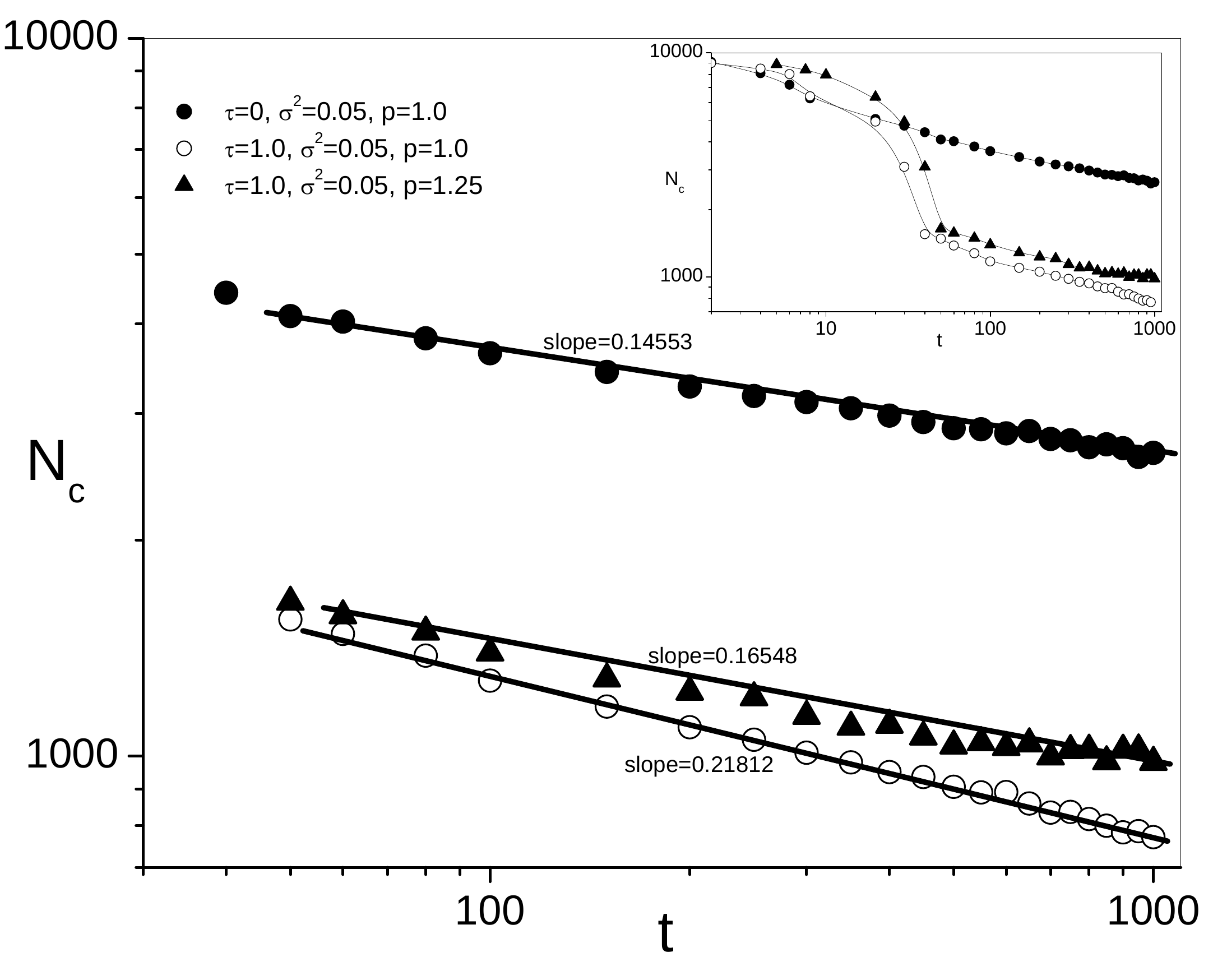} \caption{
 The plot presents the variation of the cluster
number as a function of time at long time limit for different
values of the noise correlation time ($\tau$) and the non-Gaussian
parameter(p). The other parameter sets are $\lambda=1,\;
\alpha=1,\; \rho=1\; \epsilon=0.25$. In the inset same plot which
covers both both the short time as well as long time limits.}
\end{figure}

Finally, we come to consider one of the important aspects of the
present study. How the aggregation kinetics depends on the noise
correlation time? To this aim we have investigated the aggregation
processes through a set of simulation on a system involving
$10^{4}$ particles. Our observation shows (Fig.5) that both in the
Markovian and non-Markovian limits cluster number rapidly
decreases with time at short time regime and after that it slowly
varies . However, For a detail analysis of the influence of the
noise correlation and non-Gaussian parameter on the rate of
aggregation processes we set an approximate power law of the
following form.
\begin{eqnarray}\label{3.2}
N_c(t)\sim t^{-z} \; \;.
\end{eqnarray}
This type of algebraic decay law is valid for all $\tau$ and $p$
values provided $t$ is long enough (see the inset of the Fig.5).
From Fig.5 it is surprising to note that the exponent $z$ in the
power law increases more than {\it fifty percent} compared to
Markovian case for $\tau=1.0$. Thus it puts further evidence to
consider that the noise correlation time enhances the cooperative
effect in the aggregation dynamics through modification of drift
term as well as reducing the noise variance for the large noise
strength case. Before going to leave this section we mention that
the exponent $z$ is little bit smaller for colored non Gaussian
noise  compared to Gaussian colored noise as a result of higher
noise variance for former than later for the given noise strength.

\section{Conclusion}
Based on the numerical simulation of stochastic dynamics
associated with colored non-Gaussian noise we have studied the
aggregation kinetics of an self-induced model. Such type of
self-induced model is used to deals with a certain class of
problems in social biological problem. In this modelled, each
individuals represented by a Brownian particle, which undergoes a
drift velocity depending upon the population unbalance perceived
by a single individual between left and right. Through a set of
numerical simulation we have examined the effect of the noise
correlation and non-Gaussian character of the noise in the
self-induced aggregation dynamics. Our main observation includes
the following points:

(i) The colored noise can induce aggregation even at large noise
strength (Fig.1). The aggregation may disappear if one switches
from Gaussian to non Gaussian noise.

(ii) The critical noise strength rapidly increases with noise
correlation for non Gaussian noise than the non gaussian one.

(iii) The cluster formation kinetics follows a power law for the
variation of cluster number with time like, $N_c(t)\sim t^{-z}$ at
long time. The exponent $z$ remarkable increases for non Markovian
case.

We hope that our observations will be useful to understand the
aggregation processes in the social biological system.

\acknowledgments PKG is thankful to the Council of Scientific and
Industrial Research, Government of India, for a fellowship.

\end{document}